\documentclass[11pt,twoside]{article}
\usepackage{asp2014}

\aspSuppressVolSlug
\resetcounters
\begin{document}

\title{DASCH: Bringing 100+ Years of Photographic Data into the 21st Century and Beyond}

\author{Peter K.\ G.\ Williams$^{1}$}
\affil{$^1$Center for Astrophysics | Harvard \& Smithsonian, Cambridge, MA, USA; \email{pwilliams@cfa.harvard.edu}}
\paperauthor{Peter~K.~G.~Williams}{pwilliams@cfa.harvard.edu}{0000-0003-3734-3587}{Center for Astrophysics | Harvard \& Smithsonian}{}{Cambridge}{MA}{02138}{USA}

% Author index entries
%\aindex{Williams,~P.}

\begin{abstract}
The Harvard College Observatory was the preeminent astronomical data center of
the early 20th century: it gathered and archived an enormous collection of glass
photographic plates that became, and remains, the largest in the world. For
nearly twenty years DASCH (Digital Access to a Sky Century @ Harvard) actively
digitized this library using a one-of-a kind plate scanner. In early 2024, after
470,000 scans, the DASCH project finished. Now, this unique analog dataset can
be integrated into 21st-century, digital analyses. The key DASCH data products
include $\sim$200 TB of plate images, $\sim$16 TB of calibrated light curves,
and a variety of supporting metadata and calibration outputs. Virtually every
part of the sky is covered by thousands of DASCH images with a time baseline
spanning more than 100 years; most stars brighter than $B \sim 15$ have hundreds
or thousands of detections. DASCH Data Release 7, issued in late 2024,
represents the culmination of the DASCH scanning project.
\end{abstract}

\section{Introduction}

Harvard College Observatory (HCO) holds the world's largest collection of
astronomical photographic glass plates. About 550,000 of these plates are stored
along with a variety of other artifacts in three stories of the HCO ``plate
stacks'' (formally, the HCO Astronomical Photographic Glass Plate Collection) at
60 Garden St., now home to the Center for Astrophysics | Harvard \& Smithsonian
(CfA). The history of this collection dates back to the 1880's, when Anna Palmer
Draper began funding astronomical work at the university in memory of her late
husband, Henry Draper. The most recent plate in the collection was exposed on
August 13th, 1992. The HCO plates therefore span the entire history of
photographic astronomy: from the first astronomical uses of dry gelatin plates
to the ultimate triumph of the CCD.

The Harvard plate collection remains relevant to modern research primarily
because of this long time baseline. In aggregate, the plates provide access to
time-domain phenomena at a range of timescales an order of magnitude longer and
larger than can be probed through purely born-digital data. The
ten-to-hundred-year timescales probed by the Harvard plates are relevant to
diverse astrophysical phenomena such as stellar magnetic activity, outbursts
from galactic black hole transients, supernovae in galaxy clusters, or quasar
flares \citep[see, e.g.,][]{gtls12, wbl+24}.

Motivated by this potential, the DASCH project \citep[Digital Access to a Sky
Century @ Harvard;][]{the.dasch} was launched in the early 2000's by Prof.\ Josh
Grindlay at HCO. DASCH aimed to digitize the Harvard plate collection and unlock
its potential for modern astrophysical research. This effort started with the
development of a custom high-speed plate scanner \citep{sgl+06}, which had its
official first light on July 26, 2005. Paired with this undertaking was the
development of a sophisticated pipeline capable of determining science-grade
astrometric and photometric calibrations of the resulting plate images, as well
as database infrastructure to assemble the resulting measurements into
lightcurves and make them available \citep{ltg+10,lgt+11,slg+11,tgls13}.

\section{Completion of the DASCH Project}

In the time since the major components of the DASCH data processing pipeline
were put into place, project effort has focused on the monumental task of
scanning the $\sim$430,000 Harvard plates judged to be useful for time-domain
photometry. Plates not selected for DASCH scanning include spectrum plates; ones
with gratings or prisms in the optical path; ones with superimposed grids; ones
too damaged to handle; ones with no visible stars; circular plates in the
``ADH'' series; as well as the inevitable plates that turned out to be missing
or mislabeled.

The scanning effort suffered from three major setbacks. First, on January 17,
2016, a city water main on the CfA campus burst, flooding the basement floor of
the plate stacks and submerging around 61,000 plates along with many elements of
the DASCH scanning system in 38 inches of water \citep{g17}. Thanks to quick
thinking and action across the CfA community, no plates were lost: the affected
items were promptly taken outside to freeze them in order to prevent the growth
of mold. After the initial emergency was resolved, methods to clean the plates
were developed in conjunction with the Harvard Weissman Preservation Center, and
the scanner was repaired and returned to service. Scanning resumed 234 days
after the flood.

Second, the outbreak of the COVID-19 in 2020 affected the DASCH project no less
than it affected any other aspect of life. Scanning shut down for 125 days, and
then resumed with social-distancing protocols in place.

Third, on May 4, 2021, it was discovered that two DASCH servers had suffered
unrecoverable failures in their RAID storage arrays; the simultaneous failures
suggest a power event at Harvard's Boston datacenter. At the time of the event,
DASCH scanning was approximately 97\% complete. Contemporaneous turnover in
project staff led to a hiatus in DASCH operations. The present author joined the
project in January 2023 and began work to restore failed systems, reconstruct
operational procedures, and modernize and improve the data processing pipeline.
Regular scanning restarted on November 9, 2023 (a downtime of 918 days). DASCH
scanning formally completed on March 28, 2024, closing out nearly two decades of
effort.

\section{DASCH Data Release 7}

The newest DASCH data release, Data Release 7 (DR7), was issued on December 29,
2024\footnote{\url{https://dasch.cfa.harvard.edu/dr7/}}. DR7 represents the
culmination of the DASCH scanning project.

The central DASCH data products are FITS-format images of each plate. These are
referred to as ``mosaics'' because they are constructed from many individual
exposures of the DASCH scanner camera. DR7 includes mosaics of 429,274 plates
with an average data size of $\sim$750 megabytes. About 97\% of the mosaics have
astrometric calibrations, and about 89\% of them have photometric calibrations.

Most DASCH users, however, will interact chiefly with the database of lightcurve
data derived from the DASCH plate images. The principal DASCH lightcurve data
product is a database of 23,574,404,199 Johnson $B$ magnitudes of 252,458,490
sources anchored to a reference catalog that is a combination of version 2.3.2
of the Hubble \emph{Guide Star Catalog} \citep{the.gsc2} and Data Release 8 of
the AAVSO Photometric All-Sky Survey (APASS) catalog \citep{the.apass,l12c}. An
alternative lightcurve database is based on the ATLAS-refcat2 catalog
\citep{the.atlasrefcat2}. While this catalog has superior completeness and
astrometry (being anchored to \emph{Gaia} DR2) its photometric calibration is to
the SDSS $g$ system, which only somewhat matches the response of the blue
photographic emulsions used by the vast majority of the Harvard plates. This
induces systematics that can include false long-term trends due to changes in
the emulsions used over time. \emph{The ATLAS lightcurves should only be used
with a great deal of caution.} Supporting these datasets are a number of
additional products, including $\sim$800,000 photographs of the plates and their
paper jackets; $\sim$166,000 photographs of the observing logbooks documenting
the plates and selected historical astronomer notebooks discussing them;
databases of information regarding the HCO plate stacks holdings; raw data and
calibrations from the DASCH scanner; the source code to the entire DASCH
software stack; and logs related to the DR7 data processing.

Finally, DR7 itself is described by a ``digital inventory'' that documents every
single digital file associated with the DASCH project that was judged worthy of
long-term preservation. The archive consists of 33,791,530 files totaling
745,627,062,858,355 bytes ($\sim$678 TiB) of data. The inventory documents the
size and MD5 digest of every file, as well as at least one storage location on
the Amazon S3 service, as well as potential additional storage locations on
magnetic tapes held at HCO. While the DASCH archive itself is too large to hold
at present-day research data repositories, the inventory has been deposited at
Zenodo \citep{w24}.

The DASCH data access services and documentation were completely rebuilt for
DR7. The recommended method for obtaining and analyzing DASCH data is now a
Python package called \emph{daschlab}. See the DR7 website for documentation.

\section{Remarks}

The DASCH project was the work of hundreds of people, and it in turn built on
the legacy of a century of painstaking labor at Harvard College Observatory.
Above all, it should be remembered that the Harvard plates are inextricably
linked with the pioneering ``women astronomical computers'' who curated and
analyzed the collection \citep{s16}. The plates should therefore be thought of
as artifacts with historical and cultural, not just scientific, importance, and
treated with according respect. This combination of historical meaning and
cutting-edge scientific relevance is a perhaps unique legacy of the DASCH
project.

\acknowledgments

PKGW thanks Daina Bouquin for securing the funding that made DASCH Data Release
7 possible, as well as the entire staff of the Wolbach Library for their role in
ensuring the project's successful completion. He also thanks every member of the
2023--2024 Plate Stacks team and acknowledges their many predecessors. The DASCH
website attempts to name everyone who was directly involved in the project.

The DASCH project at Harvard is grateful for partial support from NSF grants
AST-0407380, AST-0909073, and AST-1313370; which should be acknowledged in all
papers making use of DASCH data. Work on DASCH Data Release 7 received support
from the Smithsonian American Women's History Initiative Pool. We acknowledge
the one-time gift of the Cornel and Cynthia K. Sarosdy Fund for DASCH.

\bibliography{C303}

% if we have space left, we might add a conference photograph here. Leave commented for now.
% \bookpartphoto[width=1.0\textwidth]{foobar.eps}{FooBar Photo (Photo: Any Photographer)}

\end{document}